\documentclass[iop,apjl,twocolappendix]{emulateapj}
\usepackage{ulem,amsmath,amssymb,amstext,apjfonts,xcolor}%
\usepackage[backref,breaklinks,colorlinks,citecolor=blue,linkcolor=magenta,urlcolor=blue]{hyperref}

\begin{document}

\title{Dependence of Solar Wind Proton Temperature on the Polarisation Properties\\of Alfv\'enic Fluctuations at Ion-kinetic Scales}

\author{L. D. Woodham,$^{*,1}$ R. T. Wicks,$^{2,3,4}$ D. Verscharen,$^{3,5}$ J. M. TenBarge,$^{6,7}$ \& G. G. Howes$^{8}$}

\email{*email: l.woodham@imperial.ac.uk}

\affiliation{\\$^{1}$Department of Physics, The Blackett Laboratory, Imperial College London, London, SW7 2AZ, UK\\$^{2}$Department of Mathematics, Physics \& Electrical Engineering, Northumbria University, Newcastle upon Tyne, NE1 8ST, UK\\$^{3}$Mullard Space Science Laboratory, University College London, Holmbury St. Mary, Surrey RH5 6NT, UK\\$^{4}$Institute of Risk and Disaster Reduction, University College London, London, WC1E 6BT, UK\\$^{5}$Space Science Center, University of New Hampshire, Durham, NH 03824, USA\\$^{6}$Department of Astrophysical Sciences, Princeton University, Princeton, NJ 08544, USA\\$^{7}$Princeton Plasma Physics Laboratory, Princeton, NJ 08540, USA\\$^{8}$Department of Physics and Astronomy, University of Iowa, Iowa City, IA 52242, USA}

\begin{abstract}

We use fluctuating magnetic helicity to investigate the polarisation properties of Alfv\'enic fluctuations at ion-kinetic scales in the solar wind as a function of $\beta_p$, the ratio of proton thermal pressure to magnetic pressure, and $\theta_{vB}$, the angle between the proton flow and local mean magnetic field, $\mathbf{B}_0$. Using almost 15 years of \textit{Wind} observations, we separate the contributions to helicity from fluctuations with wave-vectors, $\textbf{k}$, quasi-parallel and oblique to $\mathbf{B}_0$, finding that the helicity of Alfv\'enic fluctuations is consistent with predictions from linear Vlasov theory. This result suggests that the non-linear turbulent fluctuations at these scales share at least some polarisation properties with Alfv\'en waves. We also investigate the dependence of proton temperature in the $\beta_p$-$\theta_{vB}$ plane to probe for possible signatures of turbulent dissipation, finding that it correlates with $\theta_{vB}$. The proton temperature parallel to $\mathbf{B}_0$ is higher in the parameter space where we measure the helicity of right-handed Alfv\'enic fluctuations, and the temperature perpendicular to $\mathbf{B}_0$ is higher where we measure left-handed fluctuations. This finding is inconsistent with the general assumption that by sampling different $\theta_{vB}$ in the solar wind we can analyse the dependence of the turbulence distribution on $\theta_{kB}$, the angle between $\textbf{k}$ and $\mathbf{B}_0$. After ruling out both instrumental and expansion effects, we conclude that our results provide new evidence for the importance of local kinetic processes that depend on $\theta_{vB}$ in determining proton temperature in the solar wind.

\end{abstract}

%\maketitle

\section{Introduction} \label{sec:1}

The solar wind is a variable flow of plasma that escapes from the solar corona out into the heliosphere. In-situ measurements of the solar wind provide insights into the fundamental physical processes occurring in expanding astrophysical plasmas. Fluctuations in the solar wind plasma and electromagnetic fields exist over many orders of magnitude in scale, linking both microscopic and macroscopic processes \citep[see][and references therein]{Matteini2012,Alexandrova2013}. The couplings between large-scale dynamics and small-scale kinetic processes are central to our understanding of energy transport and heating in these plasmas \citep{Verscharen2019}. There are still many open questions in regards to wave dissipation and plasma heating in collisionless plasmas. Understanding these mechanisms in the collisionless solar wind plasma is a major outstanding problem in the field of heliophysics research.

In solar wind originating from open field lines in the corona, fluctuations are predominantly Alfv\'enic \citep{ColemanJr.1968,Belcher1969,Belcher1971}, with only a small compressional component \citep{Howes2012,Klein2012,Chen2016a,Safrankova2019}. At scales $10^5\lesssim L\lesssim10^8$ m, called the inertial range, non-linear interactions between fluctuations lead to a turbulent cascade of energy towards smaller scales \citep{Tu1995,Bruno2013}. This range is characterised by fluctuations with increasing anisotropy toward smaller scales, $k_\perp\gg k_\|$, where $k_\|$ and $k_\perp$ are components of the wave-vector, $\textbf{k}$, in the direction parallel and perpendicular to the local mean magnetic field, $\mathbf{B}_0$, respectively \citep{Horbury2008,MacBride2010,Wicks2010a,Chen2011,Chen2012a}. At scales close to the proton inertial length, $d_p$, and proton gyro-radius, $\rho_p$,  typically $L\sim10^5$ m at 1 au, the properties of the fluctuations change due to Hall \citep{Galtier2006,Galtier2007} and finite-Larmor-radius \citep{Howes2006,Schekochihin2009,Boldyrev2012} effects. The non-linear turbulent fluctuations at these ion-kinetic scales exhibit some properties that are consistent with those of kinetic Alfvén waves \citep[KAWs;][]{Leamon1999,Bale2005,Howes2008b,Sahraoui2010,Woodham2019}. 

Solar wind proton velocity distribution functions (VDFs) typically deviate from local thermal equilibrium due to a low rate of collisional relaxation \citep{Kasper2008,Marsch2012,Maruca2013a,Kasper2017}. The coupling of small-scale electromagnetic fluctuations and the kinetic features of the proton VDFs can lead to energy transfer between fluctuating fields and the particles. Collisionless damping of these fluctuations can lead to dissipation of turbulence via wave-particle interactions such as Landau \citep{Leamon1999,Howes2008b} and cyclotron \citep{Marsch1982,Marsch2003a,Isenberg} resonance, or other processes such as stochastic heating \citep{Chandran2010,Chandran2013} and reconnection-based mechanisms \citep{Sundkvist2007,Perri2012}. These mechanisms are dependent on the modes present and the background plasma conditions, i.e., a function of the ratio of proton thermal pressure to magnetic pressure, $\beta_p=n_pk_BT_{p}/(B_0^2/2\mu_0)$, where $n_p$ is the proton density, and $T_p$ is the proton temperature. Each mechanism leads to distinct fine structure in proton VDFs, increasing the effective collision rate. These processes ultimately lead to plasma heating, and therefore, changes in the macroscopic properties of the plasma \citep[e.g.,][]{Marsch2006}.

In addition to damping of turbulent fluctuations, non-Maxwellian features of solar wind VDFs such as temperature anisotropies relative to $\mathbf{B}_0$, beam populations, and relative drifts between plasma species provide sources of free energy for instabilities at ion-kinetic scales \citep{Kasper2002a,Kasper2008,Kasper2013,Hellinger2006,Bale2009,Maruca2012,Bourouaine2013,Gary2015a,Kasper2018}. These modes grow until the free energy source is removed, acting to limit departure from an isotropic Maxwellian. Ion-scale kinetic instabilities are prevalent in collisionally young solar wind \citep{Klein2018,Klein2019}, although the interaction between instabilities and the background turbulence is still poorly understood \citep[e.g.,][]{Klein2015}. As the solar wind flows out into the heliosphere, instabilities, local heating, heat flux, and collisions all alter the macroscopic thermodynamics of the plasma through coupling between small-scale local processes and large-scale dynamics. These processes lead to a deviation from Chew-Goldberger-Low theory \citep[CGL;][]{Chew1956} for double adiabatic expansion \citep{Matteini2007}.

Alfv\'enic fluctuations in the solar wind are characterised by magnetic field fluctuations, $\delta \textbf{B}$, with a quasi-constant field magnitude, $|\textbf{B}|$. Since the fluctuations have large amplitudes, $\delta \textbf{B}/\textbf{B}_0\sim1$, the magnetic field vector traces out a sphere of constant radius \citep{Barnes1981}, leading to fluctuations in the angle, $\theta_{RB}$, between the local field, $\textbf{B}=\textbf{B}_0+\delta\textbf{B}$, and the radial direction. These fluctuations correlate with proton motion and therefore, the solar wind bulk velocity, $\textbf{v}_{sw}$, also exhibits a dependence on $\theta_{RB}$ \citep{Matteini2014,Matteini2015}. If these fluctuations play a role in plasma heating, we also expect a correlation between them and the proton temperature. Recent studies have shown that the proton temperature anisotropy at 1 au exhibits a dependence on $\theta_{RB}$ \citep{DAmicis2019a} that is not present closer to the Sun \citep{Horbury2018}, suggesting ongoing dynamical processes related to these fluctuations in the solar wind. In fact, larger wave power in transverse Alfv\'enic fluctuations is also correlated with proton temperature anisotropy \citep{Bourouaine2010}, consistent with an increase in fluctuations in $\theta_{RB}$.

Single-spacecraft observations have an inherent spatio-temporal ambiguity that complicates investigation of the coupling between Alfv\'enic fluctuations and the plasma bulk parameters. These measurements are restricted to the sampling of a time series defined by the trajectory of the spacecraft with respect to the flow velocity, $\textbf{v}_{sw}$. This limitation means that we can only resolve the component of $\textbf{k}$ along the sampling direction, i.e., predominantly the radial direction. Previous studies \citep[e.g.,][]{Horbury2008,Wicks2010a,He2011,Podesta2011} assume that the underlying distribution of turbulence in the solar wind is independent of $\theta_{vB}$, the angle between $\textbf{v}_{sw}$ and $\textbf{B}_0$. Based on this assumption, these studies use measurements of the solar wind plasma at different $\theta_{vB}$ to probe the turbulence as a function of $\theta_{kB}$, the angle between $\textbf{k}$ and $\textbf{B}_0$. However, if there is indeed a dependence of the plasma bulk parameters (including the temperature and temperature anisotropies as observed) on $\theta_{RB}\simeq\theta_{vB}$, then this assumption may not be valid.

In this paper, we investigate whether the solar wind proton temperature anisotropy depends on the polarisation properties of small-scale Alfv\'enic fluctuations, and hence $\theta_{vB}$, in the context of turbulent dissipation. In Section \ref{sec:2}, we discuss the linear theory and polarisation properties of Alfv\'en waves. In Sections \ref{sec:3} and \ref{sec:Method}, we describe our analysis methods, using single-spacecraft measurements to measure the polarisation properties of Alfv\'enic fluctuations at ion-kinetic scales in the solar wind. We present our main results in Section \ref{sec:Results}, testing how the dissipation of turbulence at these scales affects the macroscopic bulk properties of the solar wind. We show that Alfv\'enic fluctuations present at ion-kinetic scales in the solar wind share at least some polarisation properties with Alfv\'en waves from linear Vlasov theory. By also investigating the statistical distribution of proton temperature in the $\beta_p$-$\theta_{vB}$ plane, we find that there is a clear dependence in this reduced parameter space that also correlates with the magnetic helicity of Alfv\'enic fluctuations. We discuss the implications of our results in Section \ref{sec:Discussion}, namely that we cannot sample different $\theta_{vB}$ to analyse the dependence of the turbulence on $\theta_{kB}$ without considering other plasma properties. In Section \ref{sec:Caveats}, we consider both instrumental and expansion effects, showing that they do not account for the observed temperature distribution. Finally, in Section \ref{sec:Conclusions}, we conclude that our results provide new evidence for the importance of local kinetic processes that depend on $\theta_{vB}$ in determining proton temperature in the solar wind.

\section{Polarisation Properties of Alfv\'en Waves} \label{sec:2}

In collisionless space plasmas such as the solar wind, the linearised Vlasov equation describes linear waves and instabilities. Non-trivial solutions exist only when the complex frequency, $\omega=\omega_r+i\gamma$, solves the hot-plasma dispersion relation \citep{Stix1992}. Here, $\omega_r$ is the wave frequency and $\gamma$ is the wave growth ($\gamma>0$) or damping ($\gamma<0$) rate. One such solution is the Alfv\'en wave, which is ubiquitous in space plasmas. At $k_\|d_p\ll1$ and $k_\perp\rho_p\ll1$, this wave is incompressible and propagates along $\textbf{B}_0$ at the Alfv\'en speed, $v_A$, resulting in transverse perturbations to the field \citep{Alfven1942}. The fluctuations in velocity, $\delta\textbf{v}$, and the magnetic field, $\delta\textbf{b}$, exhibit a characteristic (anti-)correlation, $\delta\textbf{v}=\mp\delta\textbf{b}$, for propagation (parallel) anti-parallel to $\textbf{B}_0$. Here, $\textbf{b}$ is the magnetic field in Alfv\'en units, $\textbf{b}=\textbf{B}/\sqrt{\mu_0\rho}$, where $\rho$ is the plasma mass density. The Alfv\'en wave has the dispersion relation:

\begin{equation} \label{eq:AW}
	\omega_r(k)=k v_A \cos{\theta_{kB}}.
\end{equation}

\noindent Approaching ion-kinetic scales, $k_\|d_p\simeq1$ or $k_\perp\rho_p\simeq1$, the dispersion relation splits into two branches: the Alfv\'en ion-cyclotron (AIC) wave for small $\theta_{kB}$ \citep{Gary2004} and the KAW for large $\theta_{kB}$ \citep{Gary2004a}.

We define the polarisation of a wave as:

\begin{equation}
	P=-\frac{i\delta E_y}{\delta E_x}\frac{\omega_r}{\left|\omega_r\right|},
\end{equation}

\noindent where $\delta E_x$ and $\delta E_y$ are components of the Fourier amplitudes of the fluctuating electric field transverse to $\textbf{B}_0=B_0 \,\hat{\textbf{z}}$ \citep{Stix1992,Gary1993}. Therefore, $P$ gives the sense and degree of rotation in time of a fluctuating electric field vector at a fixed point in space, viewed in the direction parallel to $\textbf{B}_0$. A circularly polarised wave has $P=\pm1$, where +1 (-1) designates right-handed (left-handed) polarisation. In this definition, a right-hand polarised wave has electric field vectors that rotate in the same sense as the gyration of an electron, and a left-hand polarised wave, the same sense as ions. For more general elliptical polarisation, we take the real part, $\mathrm{Re}(P)$.

Magnetic helicity is a measure of the degree and sense of spatial rotation of the magnetic field \citep{Woltjer1958a,Woltjer1958}. It is an invariant of ideal magnetohydrodynamics (MHD) and defined as a volume integral over all space:

\begin{equation}
	H_{m} \equiv \int_V \mathbf{A} \cdot \mathbf{B} \, d^{3} \mathbf{r},
\end{equation}

\noindent where $\textbf{A}$ is the magnetic vector potential defined by $\textbf{B}=\nabla\times\textbf{A}$. \citet{Matthaeus1982a} propose the fluctuating magnetic helicity, $H_m'(\textbf{k})$, as a diagnostic of solar wind fluctuations, which in spectral form (i.e., in Fourier space) is defined as: 

\begin{equation}
	H_m'(\mathbf{k})\equiv\delta\mathbf{A}(\mathbf{k})\cdot\delta\mathbf{B}^*(\mathbf{k}),
\end{equation}

\noindent where $\delta\mathbf{A}$ is the fluctuating vector potential, and the asterisk indicates the complex conjugate of the Fourier coefficients \citep{Matthaeus1982b}. This definition removes contributions to the helicity arising from $\textbf{B}_0$. By assuming the Coulomb gauge, $\nabla\cdot\mathbf{A}=0$, the fluctuating magnetic helicity can be written:

\begin{equation} \label{eq:NRhel}
	H_{m}'(\mathbf{k})=i \frac{\delta B_{y} \delta  B_{z}^{*}-\delta B_{y}^{*} \delta  B_{z}}{k_{x}},
\end{equation}

\noindent where the components of $\delta \mathbf{B}(\mathbf{k})$ are Fourier coefficients of a wave mode with $\mathbf{k}$. This result is invariant under cyclic permutations of the three components $x,y,z$ \citep[See Equation 2 in][]{Howes2010}. We define the normalised fluctuating magnetic helicity density as:

\begin{equation} \label{eq:normhel}
	\sigma_m(\textbf{k})\equiv\frac{kH_m'(\textbf{k})}{\left|\delta\mathbf{B}(\textbf{k})\right|^2},
\end{equation}

\noindent where $\left|\delta\mathbf{B}(\textbf{k})\right|^2=\delta B_{x}^{*} \delta  B_{x}+\delta B_{y}^{*} \delta  B_{y}+\delta B_{z}^{*} \delta  B_{z}$. Here, $\sigma_m(\textbf{k})$ is dimensionless and takes values in the interval $[-1,1]$, where $\sigma_m=-1$ indicates fluctuations with purely left-handed helicity, and $\sigma_m=+1$ purely right-handed helicity. A value of $\sigma_m=0$ indicates no overall coherence, i.e., there are either no fluctuations with coherent handedness or there is equal power in both left-handed and right-handed components so that the net value is zero. 

\citet{Gary1986} first explored the dependence of $\textrm{Re}(P)$ for small-scale Alfv\'en waves on different parameters by numerically solving the full electromagnetic dispersion relation, showing that it changes sign depending on both $\theta_{kB}$ and $\beta_p$. In the cold-plasma limit ($\beta_p\ll1$), the Alfv\'en wave has $\textrm{Re}(P)<0$ for all $\theta_{kB}$. However, from linear Vlasov theory, at $\beta_p\simeq10^{-2}$, the wave has $\textrm{Re}(P)<0$ for $0^\circ \leq \theta_{kB}\lesssim 80^\circ$, but has $\textrm{Re}(P)>0$ for $\theta_{kB}\gtrsim80^\circ$. As $\beta_p$ increases, the wave has $\textrm{Re}(P)>0$ for an increasing range of oblique angles so that at $\beta_p\simeq10$, the transition occurs at about $40^\circ$. This result reveals that the changing polarisation properties on both $\theta_{kB}$ and $\beta_p$ will affect possible wave-particle interactions, and hence turbulence damping mechanisms that can occur in a plasma. For example, left-handed AIC waves can cyclotron resonate with ions, leading to heating perpendicular to $\textbf{B}_0$. On the other hand, right-handed KAWs are compressive at small scales, giving rise to density fluctuations and a non-zero component of the wave electric field, $E_\|\neq0$. Hence, KAWs can Landau resonate with both electrons and ions, leading to heating parallel to $\textbf{B}_0$.

\begin{figure}
	\centering
	\includegraphics[width=\linewidth]{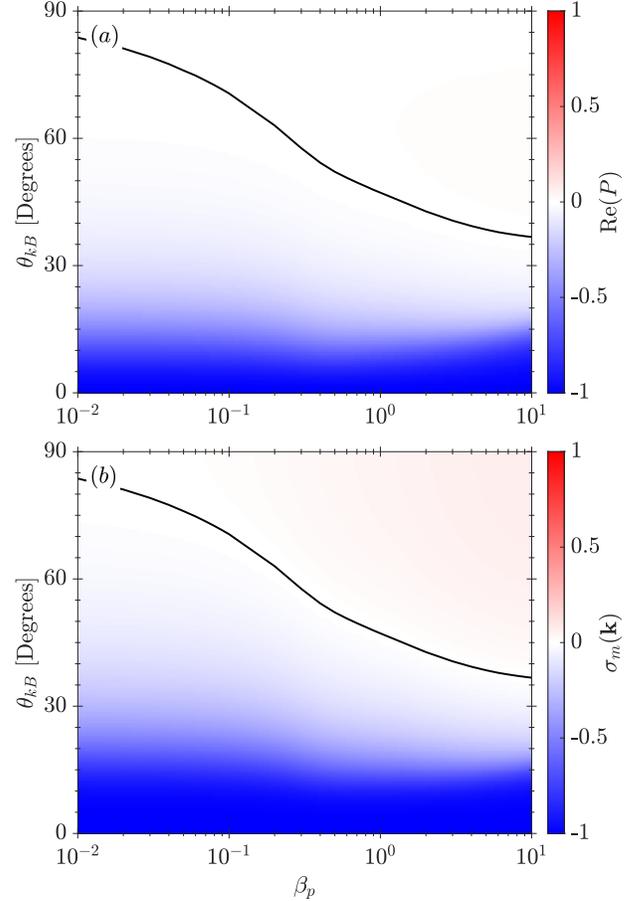} 
	\caption{$(a)$ The real part of the polarisation, $\textrm{Re}(P)$, and $(b)$ normalised fluctuating magnetic helicity density, $\sigma_m(\textbf{k})$, of Alfv\'en waves with $kd_p=0.05$ as a function of $\beta_p$ and $\theta_{kB}$, calculated using the NHDS code (see main text). The black lines indicate the isocontours $\textrm{Re}(P)=0$ and $\sigma_m(\textbf{k})=0$.}
	\label{fig:1}
\end{figure}

We plot both $\textrm{Re}(P)$ and $\sigma_m(\textbf{k})$ for Alfv\'enic fluctuations across the $\beta_p$-$\theta_{kB}$ plane in Figure \ref{fig:1}. The black lines are isocontours of  $\textrm{Re}(P)=0$ and $\sigma_m(\textbf{k})=0$, respectively. We note that for waves with $\theta_{kB}\simeq0$, there is no difference between the values of $\textrm{Re}(P)$ and $\sigma_m(\textbf{k})$. To calculate these lines, we solve the linear Vlasov equation using the New Hampshire Dispersion relation Solver \citep[NHDS;][]{Verscharen2013a,Verscharen2018}. Here, $\mathbf{k}=k_{\perp }\hat{\textbf{x}}+k_{\|}\hat{\textbf{z}}$, and we assume a plasma consisting of protons and electrons with isotropic Maxwellian distributions, equal density and temperature, and no drifting components. We set $k d_p=0.05$,\footnote{The black lines in Figure \ref{fig:1} are constant over the range: $k d_p=[0.01,1]$.} where the angle $\theta_{kB}$ defines $k_\perp=k\sin{\theta_{kB}}$ and $k_\|=k\cos{\theta_{kB}}$. Therefore, $k_\|d_p$ and $k_\perp\rho_p$ change throughout the $\beta_p$-$\theta_{kB}$ plane,\footnote{The scales $d_p$ and $\rho_p$ are related by: $\rho_p=d_p\sqrt{\beta_p}$.} while the normalised scale of the waves remains constant. We also set $v_A/c=10^{-4}$, which is typical for solar wind conditions where $v_A\simeq50$ km/s \citep{Klein2019}. While our assumption of an isotropic proton-electron plasma is not truly representative of the more complex ion VDFs typically observed in the solar wind, protons remain the most important ion component for solar wind interaction with Alfv\'enic fluctuations. Therefore, we expect that the polarisation properties of Alfv\'enic fluctuations in the solar wind are adequately described by the theoretical description provided in Figure \ref{fig:1}.

\section{Reduced Spectra from Spacecraft Measurements} \label{sec:3}

In the solar wind, the polarisation properties of fluctuations are typically determined using the fluctuating magnetic helicity. However, from a single-spacecraft time series of magnetic field measurements, it is only possible to determine a reduced form of the helicity \citep{Batchelor1970,Montgomery1981,Matthaeus1982a}:

\begin{equation}
	H'^r_m(\omega_{sc})=\frac{2\,\mathrm{Im}\left\lbrace\mathcal{P}^r_{TN}(\omega_{sc})\right\rbrace}{k_r},
	\label{eq:redflucthel}
\end{equation}

\noindent where $\omega_{sc}$ is the frequency of the fluctuations in the spacecraft frame, $k_r=k\cos{\theta_{kv}}$ is the component of the wave-vector along the flow direction of the solar wind plasma, $\textbf{v}_{sw}$, and $\theta_{kv}$ is the angle between $\textbf{k}$ and $\textbf{v}_{sw}$. Here,

\begin{equation}
	\mathcal{P}^r_{ij}(\omega_{sc})=\delta B_i^*(\omega_{sc}) \cdot \delta B_j(\omega_{sc})
	\label{eq:RPST}
\end{equation}

\noindent is the reduced power spectral tensor, where the $\delta B_i(\omega_{sc})$ are the complex Fourier coefficients of the time series of $\mathbf{B}$ in radial-tangential-normal (RTN) coordinates.\footnote{In the RTN coordinate system, $\hat{\mathbf{R}}$ is the unit vector from the Sun towards the spacecraft, $\hat{\mathbf{T}}$ is the cross-product of the solar rotation axis and $\hat{\mathbf{R}}$, and $\hat{\mathbf{N}}$ completes the right-handed triad.} In this coordinate system, the solar wind flow is approximately radial, $\textbf{v}_{sw}\simeq v_{sw}\,\hat{\mathbf{R}}$. The reduced tensor is an integral of the true spectral tensor, $\mathcal{P}_{ij}(\textbf{k})$ \citep{Fredricks1976,Forman2011,Wicks2012}:

\begin{equation} \label{eq:Reduced}
	\mathcal{P}_{i j}^{r}\left(\omega_{sc}\right)=\int \mathcal{P}_{i j}(\mathbf{k}) \, \delta\left[\omega_{sc}-\left(\mathbf{k} \cdot \mathbf{v}_{s w}+\omega_{pl}\right)\right] \, d^{3} \mathbf{k}.
\end{equation}

\noindent Taylor's hypothesis \citep{Taylor1938} assumes that the fluctuations in the solar wind evolve slowly as they are advected past the spacecraft so that the plasma-frame frequency, $\omega_{pl}$, satisfies $|\omega_{pl}|\ll|\mathbf{k} \cdot \mathbf{v}_{s w}|$ \citep{Matthaeus1982b,Perri2010a}. Therefore, the Doppler shift of the fluctuations into the spacecraft frame becomes:

\begin{equation} \label{eq:Taylor}
	\omega_{sc}=\omega_{pl}+\mathbf{k} \cdot \mathbf{v}_{s w}\simeq\mathbf{k} \cdot \mathbf{v}_{s w}\equiv k_rv_{sw},
\end{equation}

\noindent so that the $\omega_{pl}$ term drops from Equation \ref{eq:Reduced}. Then, a time series of magnetic field measurements under these assumptions represents a spatial cut through the plasma and we can write $\mathcal{P}_{i j}^{r}$ and $H'^r_m$ as functions $k_r$ using Equation \ref{eq:Taylor}. However, it is not possible to determine the full wave-vector, $\textbf{k}$, or $\theta_{kB}$, from single-spacecraft measurements. Since $v_A\ll v_{sw}$, Taylor's hypothesis is usually well-satisfied for Alfv\'en waves in the solar wind with the dispersion relation given by Equation \ref{eq:AW}, as well as for the small-wavelength extensions of the Alfv\'en branch under the parameters considered here \citep[see][]{Howes2014,Klein2014a}.

Based on the definition in Equation \ref{eq:normhel}, the normalised reduced fluctuating magnetic helicity density is then defined as:

\begin{equation}
	\sigma_m^r(k_r)\equiv\frac{k_rH'^r_m(k_r)}{\left|\delta\mathbf{B}(k_r)\right|^2}=\frac{2\,\mathrm{Im}\left\lbrace\mathcal{P}^r_{TN}(k_r)\right\rbrace}{\mathrm{Tr}\left\lbrace\mathcal{P}^r(k_r)\right\rbrace},
	\label{eq:reducednormhel}
\end{equation}

\noindent where $\mathrm{Tr}\lbrace\rbrace$ denotes the trace. Previous studies \citep[e.g.,][]{Horbury2008,Wicks2010a,He2011,Podesta2011} use $\theta_{vB}$ as a measure of a specific $\theta_{kB}$ in the solar wind. For example, measurements of $\sigma_m^r(k_r)$ separated as a function of $\theta_{vB}$ show a broad right-handed signature at oblique angles and a narrow left-handed signature at quasi-parallel angles \citep{He2011,He2012,He2012a,Podesta2011,Klein2014,Bruno2015,Telloni2015}. These signatures are associated with KAW-like fluctuations from the turbulent cascade and ion-kinetic instabilities, respectively \citep{Telloni2016,Woodham2019}.

By defining the field-aligned coordinate system,

\begin{equation}
	\hat{\textbf{z}}=\frac{\mathbf{B}_0}{\left|\mathbf{B}_0\right|};
	\:\hat{\textbf{y}}=-\frac{\mathbf{v}_{sw}\times\mathbf{B}_0}{\left|\mathbf{v}_{sw}\times\mathbf{B}_0\right|};
	\:\hat{\textbf{x}}=\hat{\textbf{y}}\times\hat{\textbf{z}},
	\label{eqncoords}
\end{equation}

\noindent so that $\mathbf{v}_{sw}$ lies in the $x$-$z$ plane with angle $\theta_{vB}$ from the $\hat{\textbf{z}}$ direction \citep{Wicks2012,Woodham2019}, we can decompose $\sigma_m^r(k_r)$ into the components:

\begin{equation}
	\sigma_{ij}(k_l)=\frac{2\,\mathrm{Im}\left\lbrace\mathcal{P}_{ij}^r(k_l)\right\rbrace}{\mathrm{Tr}\left\lbrace\mathcal{P}^r(k_l)\right\rbrace},
	\label{eqnhelFA}
\end{equation}

\noindent where the indices $i,j,l=x,y,z$. We derive the following relationship between the components, $\sigma_{ij}(k_l)$, and $\sigma_m^r(k_r)$ (see Appendix): 

\begin{equation}
	\sigma_{xy}(k_z) = \sigma^r_m(k_r) \frac{k_z}{k_r},
	\label{Eq:a1}
\end{equation}

\begin{equation}
	\sigma_{xz}(k_y) = -\sigma^r_m(k_r) \frac{k_y}{k_r},
	\label{Eq:a2}
\end{equation}

\noindent and

\begin{equation}
	\sigma_{yz}(k_x) = \sigma^r_m(k_r) \frac{k_x}{k_r}.
	\label{Eq:a3}
\end{equation}

\noindent For fluctuations with $k_z \gg k_x$, i.e., $\textbf{k}$ quasi-parallel to $\textbf{B}_0$, $\sigma_{xy}(k_z)$ is the dominant contribution to $\sigma^r_m(k_r)$. Similarly, $\sigma_{yz}(k_x)$ dominates for modes with $k_x \gg k_z$, i.e., $\textbf{k}$ at oblique angles, $\theta_{kB}\gtrsim60^\circ$. As the solar wind velocity is confined to the $x$-$z$ plane, we have no information about $k_y$ from single-spacecraft measurements and $\sigma_{xz}(k_y)$ is not useful in a practical sense. From Section \ref{sec:2}, we expect that AIC waves generated by kinetic instabilities have $k_z \gg k_x$. The anisotropic Alfv\'enic turbulent cascade leads to the generation of nearly perpendicular wave-vectors with $k_x \gg k_z$. Therefore, we can separate the helicity signatures of the two kinetic scale branches of the Alfvén wave using our decomposition technique.

\section{Data Analysis}\label{sec:Method}

We analyse magnetic field data from the MFI fluxgate magnetometer \citep{Lepping1995,Koval2013} and proton data from the SWE Faraday cup \citep{Ogilvie1995,Kasper2006} instruments on-board the \textit{Wind} spacecraft from Jun 2004 to Oct 2018. For each proton measurement, we define a local mean field, $\mathbf{B}_0$, averaged over the SWE integration time ($\sim$92 s). We estimate the normalised cross-helicity \citep{Matthaeus1982} for each $\sim$92 s interval,

\begin{equation}
	\sigma_{c}=\frac{2\,(\delta \mathbf{v} \cdot \delta \mathbf{b})}{|\delta \mathbf{v}|^{2}+|\delta \mathbf{b}|^{2}},
\end{equation}

\noindent where $\delta \mathbf{b}=\mathbf{b}-\langle\mathbf{b}\rangle_{1 h}$ and $\delta \mathbf{v}=\mathbf{v}_{sw}-\langle\mathbf{v}_{sw}\rangle_{1 h}$. Here, the mean is over a one hour window centred on the instantaneous values and we assume that $\textbf{v}_{sw}\simeq\textbf{v}_p$, where $\textbf{v}_p$ is the proton bulk velocity. An averaging interval of one hour gives $\sigma_c$ for fluctuations in the inertial range. The cross-helicity, $\sigma_{c} \in[-1,1]$, is a measure of the (anti-)correlation between velocity and magnetic field fluctuations, and therefore, Alfv\'enicity \citep[e.g.,][]{DAmicis2015,DAmicis2019,Stansby2019,Perrone2020}. A value $|\sigma_c| = 1$ indicates purely unbalanced Alfv\'enic fluctuations propagating in one direction, whereas $\sigma_c = 0$ indicates either balanced (equal power in opposite directions) or a lack of Alfv\'enic fluctuations. In case of $\sigma_c = 0$, we expect no coherent value of $|\sigma_m|>0$ at ion-kinetic scales.

Similarly to \citet{Woodham2019}, we account for heliospheric sector structure in the magnetic field measurements by calculating $\sigma_c$ averaged over a running window of 12 hours. For solar wind fluctuations dominantly propagating anti-sunward, the sign of $\sigma_c$ depends only on the direction of $\textbf{B}_0$. Therefore, if $\left<\sigma_c\right>>0$, we reverse the signs of the $B_{R}$ and $B_{T}$ components for each $\sim$92 s measurement so that sunward fields are rotated anti-sunward. This procedure removes the inversion of the sign of magnetic helicity due to the direction of the large-scale magnetic field with respect to the Sun.\footnote{See Table 1 in \citet{Woodham2019}.} We transform the 11 Hz magnetic field data associated with each proton measurement into field-aligned coordinates (Equation \ref{eqncoords}) using $\textbf{B}_0$ averaged over $\sim$92 s. We then compute the continuous wavelet transform \citep{Torrence1998} using a Morlet wavelet to calculate the magnetic helicity spectra, $\sigma_{xy}$ and $\sigma_{yz}$, as functions of $f_{sc}=\omega_{sc}/2\pi$ using Equation \ref{eqnhelFA}. We average the spectra over $\sim$92 s, corresponding to the SWE measurement cadence, to ensure that the fluctuations contributing to the helicity spectra persist for at least several proton gyro-periods, $2\pi/\Omega_p$, giving a clear coherent helicity signature at ion-kinetic scales.

We estimate the amplitudes of $\sigma_{xy}$ and $\sigma_{yz}$ at ion-kinetic scales by fitting a Gaussian function to the coherent peak in each spectrum at frequencies close to the Taylor-shifted frequencies, $v_{sw}/ d_p$ and $v_{sw}/ \rho_p$ \citep[see][]{Woodham2018}. We neglect any peak at $f>f_{\textrm{noise}}$, where $f_{\textrm{noise}}$ is the frequency above which instrumental noise of the MFI magnetometer becomes significant.\footnote{See Appendix in \citet{Woodham2018}.} We also reject a spectrum if the angular deviation in $\mathbf{B}$ exceeds 15$^\circ$ during the $\sim$92 s measurement period to ensure that we measure the anisotropy of fluctuations at ion-kinetic scales with sufficient accuracy (see also Section \ref{sec:uncert}). We designate the amplitude of the peak in each $\sigma_{xy}$ spectrum as $\sigma_\parallel\equiv \max_{k_z} \sigma_{xy}(k_z)$ to diagnose the helicity of the modes with $\textbf{k}$ quasi-parallel to $\mathbf{B}_0$, and $\sigma_{yz}$ as $\sigma_\perp \equiv \max_{k_x} \sigma_{yz}(k_x)$ to diagnose the helicity of the modes with $\textbf{k}$ oblique to $\mathbf{B}_0$ (see Section \ref{sec:3} and Appendix).

\begin{figure}
	\centering
	\includegraphics[width=\linewidth]{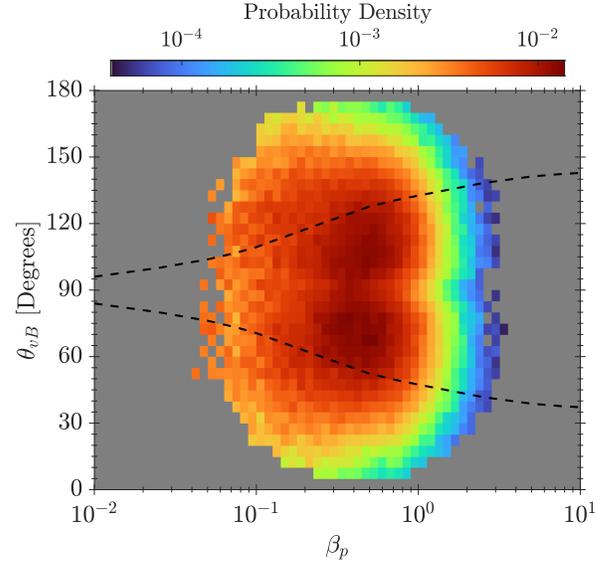} 
	\caption{Probability density distribution of solar wind data in the $\beta_p$-$\theta_{vB}$ plane, calculated using Equation \ref{eq:pdf}. The dashed black lines indicate the isocontours of $\tilde{\sigma}_m(\theta_{vB})=0$ mirrored about the line $\theta_{vB}=90^\circ$ (see main text).}
	\label{fig:2}
\end{figure}

In our analysis, we include only measurements of Alfv\'enic solar wind, $|\sigma_c|\geq0.8$, and low collisionality, $N_c<1$, which contain the strongest Alfv\'enic fluctuations with a non-zero magnetic helicity. Here, $N_c$ is the Coulomb number \citep{Maruca2013a,Kasper2017}, which estimates the number of collisional timescales for protons. We calculate $N_c$ using the proton-proton collision frequency, neglecting collisions between protons and other ions. We bin the data in $\mathrm{log}_{10}(\beta_p)$ and $\theta_{vB}$ using bins of width $\Delta \log_{10}(\beta_{p})=0.05$ and $\Delta\theta_{vB}=5^\circ$. We restrict our analysis to $0.01\leq\beta_{p}\leq10$ and include the full range of $\theta_{vB}=[0^\circ,180^\circ]$ to account for any dependence on heliospheric sector structure. In Figure \ref{fig:2}, we plot the probability density distribution of the data,

\begin{equation} \label{eq:pdf}
	\tilde{p}=\frac{n}{N\Delta\beta_{p}\Delta\theta_{vB}},
	%\left|\cos{\theta_{c}}\right|,
\end{equation}

\noindent in the $\beta_p$-$\theta_{vB}$ plane, where $n$ is the number of data points in each bin and $N$ is the total number of data points. We overplot the isocontour of $\sigma_m(\textbf{k})=0$ from Figure \ref{fig:1}$(b)$ by replacing $\theta_{kB}$ with $\theta_{vB}$, i.e., $\tilde{\sigma}_m(\theta_{vB})=0$. If we assume the turbulence is independent of $\theta_{vB}$, then any dependence on $\theta_{vB}$ exclusively reflects a dependence on $\theta_{kB}$ \citep[see][]{Horbury2008,Wicks2010a,He2011,Podesta2011}. We mirror the $\tilde{\sigma}_m(\theta_{vB})=0$ curve around the line $\theta_{vB}=90^\circ$ to account for heliospheric sector structure. The distribution of data in Figure \ref{fig:2} shows two peaks at $\theta_{vB}\sim70^\circ$ and $\theta_{vB}\sim110^\circ$ around $\beta_p\sim0.5$. There are fewer data points at quasi-parallel angles, showing that the majority of data are associated with oblique angles. Na\"ively, one would expect the distribution to follow the large-scale Parker spiral, peaking at angles $\sim45^\circ$ and $\sim135^\circ$. However, we note that $\theta_{vB}$ is calculated at $\sim$92 s timescales, over with the local mean field $\textbf{B}_0$ has already been deflected from the Parker spiral by Alfv\'enic fluctuations present at larger scales. There is also a clear $\beta_p$ dependence in Figure \ref{fig:2}, with the majority of the data lying in the range $0.1\lesssim\beta_p\lesssim1$, which is typical for quiescent solar wind \citep{WilsonIII2018}.

\section{Results} \label{sec:Results}

\begin{figure}
	\centering
	\includegraphics[width=\linewidth]{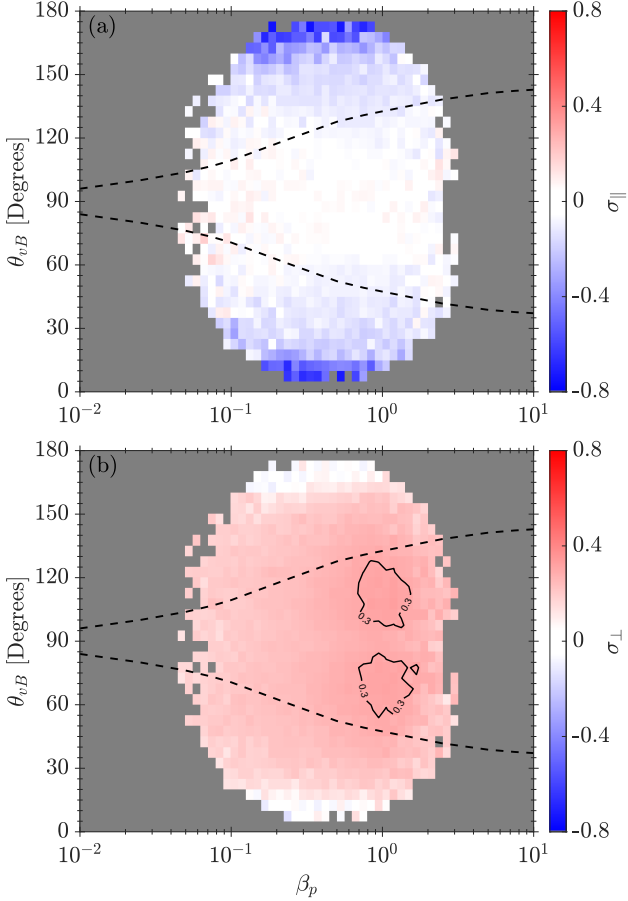} 
	\caption{$(a)$ Median $\sigma_\|$ and $(b)$ median $\sigma_\perp$ across the $\beta_{p}$-$\theta_{vB}$ plane. The dashed black lines indicate the isocontours of $\tilde{\sigma}_m(\theta_{vB})=0$ mirrored about the line $\theta_{vB}=90^\circ$. We also include contours of constant $\sigma_\perp=0.3$ in Panel $(b)$ as solid black lines.}
	\label{fig:3}
\end{figure}

In Figure \ref{fig:3}, we plot the median values of $\sigma_\|$ and $\sigma_\perp$ for each bin in the $\beta_p$-$\theta_{vB}$ plane. We neglect any bins with fewer than 20 data points to improve statistical reliability. From Figure \ref{fig:1}, we expect to measure KAW-like fluctuations with $\sigma_{\perp}>0$ in the area of the $\beta_p$-$\theta_{vB}$ plane enclosed by the two dashed lines at quasi-perpendicular angles, and AIC wave-like fluctuations with $\sigma_\|<0$ at quasi-parallel angles. Figure \ref{fig:3} is consistent with this expectation; we see a strong negative helicity signal at $0^\circ\leq\theta_{vB}\leq30^\circ$ and $150^\circ\leq\theta_{vB}\leq180^\circ$, with a minimum of $\sigma_\|\simeq-0.8$ approaching $\theta_{vB}\simeq0^\circ$, as well as a weaker positive signal of $\sigma_\perp\simeq0.4$ at angles $60^\circ\leq\theta_{vB}\leq120^\circ$. Both $\sigma_\|$ and $\sigma_\perp$ are symmetrically distributed about the line $\theta_{vB}=90^\circ$ since we remove the ambiguity in the sign of the helicity due to the direction of $\mathbf{B}_0$. The distribution of $\sigma_\|$ is consistent with the presence of quasi-parallel propagating AIC waves from kinetic instabilities in the solar wind \citep{Woodham2019,Zhao2018,Zhao2019}. Elsewhere in Figure \ref{fig:3}$(a)$, the median value of $\sigma_\|$ is zero, showing that a coherent signal of parallel-propagating fluctuations at ion-kinetic scales in the solar wind is not measured at oblique angles.

In Figure \ref{fig:3}$(b)$, there are two peaks in the median $\sigma_\perp$ close to $\beta_p\sim1$, located at $\theta_{vB}\sim70^\circ$ and $\theta_{vB}\sim110^\circ$. Despite these peaks, the signal is spread across the parameter space, except at quasi-parallel angles where $\sigma_\perp\simeq0$. We interpret this spread using Taylor's hypothesis. Due to the $\mathbf{k}\cdot\mathbf{v}_{sw}$ term in the $\delta$-function in Equation \ref{eq:Reduced}, a $\cos{\theta_{kv}}$ factor modifies the contribution of all modes to the reduced spectrum measured in the direction of $\mathbf{v}_{sw}$. If $\theta_{kv}=0^\circ$, then $\cos{\theta_{kv}}=1$, and the waves are measured at their actual $k$. However, oblique modes measured at a fixed $\omega_{sc}$ correspond to a higher $k$ in the plasma frame. Since the turbulent spectrum decreases in amplitude with increasing $k$, the reduced spectrum is most sensitive to the smallest $\mathbf{k}$ in the sampling direction. For parallel propagating fluctuations such as AIC waves, $\theta_{vB}\simeq\theta_{kv}$, but for a broader $k$-distribution of obliquely propagating fluctuations, multiple fluctuations with different $\mathbf{k}$ and therefore, different $\theta_{kB}$, contribute to a single $\theta_{vB}$ bin. The signal at $\theta_{vB}\lesssim30^\circ$ is then likely due to fluctuations with $\theta_{kB}\gtrsim60^\circ$, since they contribute to $\sigma_\perp$, i.e., have a significant $k_\perp$ component.

\begin{figure}
	\centering
	\includegraphics[width=\linewidth]{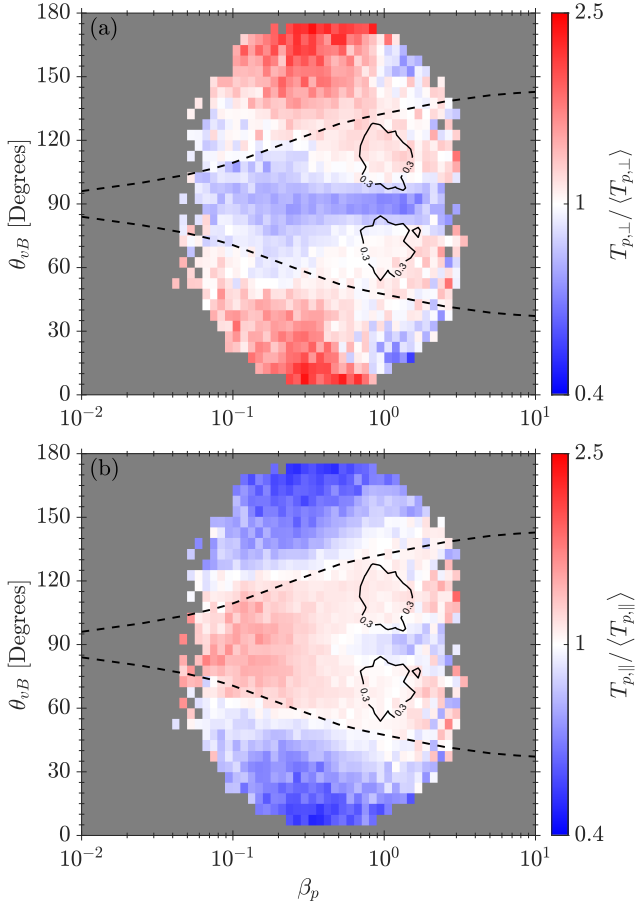} 
	\caption{$(a)$ Median proton perpendicular temperature, $T_{p,\perp}/\left<T_{p,\perp}\right>$ and $(b)$ median proton parallel temperature, $T_{p,\|}/\left<T_{p,\|}\right>$, across $\beta_{p}$-$\theta_{vB}$ space. In both panels, we column-normalise the data by the median temperature in each $\beta_p$ bin, $\left<T_{p,\perp/\|}\right>$, to remove the systematic dependency of $\beta_p$ on temperature. The dashed black lines indicate the isocontours of $\tilde{\sigma}_m(\theta_{vB})=0$ mirrored about the line $\theta_{vB}=90^\circ$. We also include contours of constant $\sigma_\perp=0.3$ from Figure \ref{fig:3}$(b)$ as black lines.}
	\label{fig:4}
\end{figure}

As the polarisation properties of small-scale Alfv\'enic fluctuations are consistent with predictions from linear theory, it is reasonable to expect that $T_p$ is also correlated in the $\beta_p$-$\theta_{vB}$ plane. This expectation follows because different Alfvénic fluctuations are associated with different dissipation mechanisms, leading to distinct heating signatures. On the other hand, if the properties of the turbulence are truly independent of $\theta_{vB}$, then we expect the dissipation mechanisms, and therefore, proton heating to be independent of $\theta_{vB}$. To test this hypothesis, we plot the median values of $T_{p,\perp}/\left<T_{p,\perp}\right>$ and $T_{p,\|}/\left<T_{p,\|}\right>$ for each bin in the $\beta_p$-$\theta_{vB}$ plane in Figure \ref{fig:4}. Here, $\left<T_{p,\perp/\|}\right>$ is the average value of $T_{p,\perp/\|}$ over all angles for each bin in $\mathrm{log}_{10}(\beta_p)$. This column normalisation removes the systematic proportionality of $T_p$ with $\beta_p$. The colour of each bin in the $\theta_{vB}$-$\beta_p$ plane, therefore, shows as a function of $\theta_{vB}$ whether the proton temperature is equal to, larger than, or smaller than the average for a specific $\beta_p$.

Figure \ref{fig:4} shows a clear dependence of the median column-normalised $T_{p,\perp}$ and $T_{p,\|}$ on both $\theta_{vB}$ and $\beta_p$. In general, we see higher $T_{p,\perp}/\left<T_{p,\perp}\right>$ at quasi-parallel angles where $\sigma_\|$ is largest in Figure \ref{fig:3}$(a)$, associated with AIC waves driven by kinetic instabilities \citep{Kasper2002b,Matteini2007,Bale2009,Maruca2012,Woodham2019}. We also see higher $T_{p,\|}/\left<T_{p,\|}\right>$ at oblique angles where $\sigma_\perp$ is largest in Figure \ref{fig:3}$(b)$, associated with KAW-like fluctuations \citep{Leamon1999,Bale2005,Howes2008b,Sahraoui2010}. However, there are also enhancements in $T_{p,\perp}/\left<T_{p,\perp}\right>$ where $\sigma_\perp\simeq0.3$, as indicated by the contours of constant $\sigma_\perp$ from Figure \ref{fig:3}$(b)$. Despite enhancements in both the column-normalised $T_{p,\perp}$ and $T_{p,\|}$ in this region of parameter space, the proton temperature remains anisotropic with $T_{p,\perp}/T_{p,\|}<1$. We note that a lack of helicity signature does not imply that waves are not present. Therefore, if the enhancements in proton temperature are associated with different dissipation mechanisms, we do not expect a perfect correlation with $\sigma_\|$ and $\sigma_\perp$ in the $\theta_{vB}$-$\beta_p$ plane.

\citet{Hellinger2014} recommend to exercise caution when bin-averaging solar wind data in a reduced parameter space. While conditional statistics have been employed in several studies \citep[e.g.,][]{Bale2009,Maruca2011,Osman2012}, this non-trivial procedure may give spurious results as a consequence of superposition of multiple correlations in the solar wind and should be interpreted cautiously. We find no evidence that the correlations shown in Figure \ref{fig:4} are caused by or related to other underlying correlations in the solar wind multi-dimensional parameter space. In particular, we rule out the known correlation between $v_{sw}$ and $T_p$ \citep{Matthaeus2006,Perrone2019} by separating our results as a function of solar wind speed, finding that Figure \ref{fig:4} is largely unchanged (not shown). This is consistent with the fact that the $\beta_p$-$\theta_{vB}$ parameter space we investigate is determined by the properties of Alfv\'enic fluctuations, which exist in both fast and slow wind \citep[e.g.,][]{DAmicis2019}.

\section{Discussion}\label{sec:Discussion}

It is well-known that Alfv\'enic turbulence is anisotropic, its properties dependent on the angle, $\theta_{kB}$. For a single spacecraft sampling in time, the common assumption of ergodicity means that we measure a statistically similar distribution of turbulent fluctuations. Hence, by sampling along different directions relative to a changing $\theta_{vB}$, we measure different components of the same distribution, e.g., the spectrum of magnetic fluctuations parallel and perpendicular to $\textbf{B}_0$. The same is true for magnetic helicity, where the left- or right-handedness is determined only by the sampling direction. Certain fluctuations may still exist and we do not measure them since we do not sample close enough to the $\textbf{k}$ of these modes for them to make a significant contribution to the $\textbf{k}\cdot\textbf{v}_{sw}$ term in Equation \ref{eq:Reduced}. Therefore, if turbulent dissipation is ongoing, we expect the resultant heating to exhibit the same distribution as the fluctuations at ion-kinetic scales. This is because the polarisation properties of solar wind fluctuations affect what dissipation mechanisms can occur.

We initially hypothesised that the proton temperature would not exhibit a systematic dependence on either $\theta_{vB}$ or $\sigma_{\|,\perp}$. However, we show a clear dependence of $T_{p,\perp/\|}/\left<T_{p,\perp/\|}\right>$ on $\theta_{vB}$ in Figure \ref{fig:4} that also correlates with the magnetic helicity signatures of different Alfv\'enic fluctuations at ion-kinetic scales. This result suggests that the properties of the turbulence also change with $\theta_{vB}$. In other words, differences in the spectra of magnetic fluctuations with changing $\theta_{vB}$ are due to both single-spacecraft sampling effects and differences in the underlying distribution of turbulent fluctuations. If this interpretation is correct, studies that sample many angles $\theta_{vB}$ as the solar wind flows past a single spacecraft to build up a picture of the turbulence in the plasma, i.e., to sample different $\theta_{kB}$, need to be interpreted very carefully \citep[e.g.,][]{Horbury2008,Wicks2010a,He2011,Podesta2011}. In this study, we measure $\theta_{vB}$ at $\sim$92 s timescales, which suppresses large-scale correlations such as the Parker-spiral. Instead, we show correlations between small-scale fluctuations with respect to a local mean field and the macroscopic proton temperature. Therefore, it is fair to assume that the dependence of $T_{p,\perp}/\left<T_{p,\perp}\right>$ and $T_{p,\|}/\left<T_{p,\|}\right>$ on $\theta_{vB}$ and $\beta_p$ reflects the differences in the localised dissipation and heating processes at ion-kinetic scales in the solar wind.

A large enough $T_{p,\perp}/T_{p,\|}$ can drive AIC waves unstable in the solar wind \citep{Kasper2002b,Matteini2007,Bale2009,Maruca2012}. The driving of these waves is enhanced by the frequent presence of an $\alpha$-particle proton differential flow or proton beam in the solar wind \citep{Podesta2011a,Podesta2011,Wicks2016,Woodham2019,Zhao2019a,Zhao2020b}. The enhancement in $T_{p,\perp}/\left<T_{p,\perp}\right>$ at quasi-parallel angles in Figure \ref{fig:4}$(a)$ is likely responsible for the driving of these modes and correlates with the peak in $\sigma_{\|}$ at these angles in Figure \ref{fig:3}$(a)$, where we measure the strongest signal. While we are unable to observe AIC waves at oblique angles using a single spacecraft, we also measure KAW-like fluctuations at these angles using $\sigma_\perp$ in Figure \ref{fig:3}$(b)$. The peaks in $\sigma_\perp$ correlate with the observed enhancement in $T_{p,\perp}/\left<T_{p,\perp}\right>$, and therefore, are consistent with the dissipation of these fluctuations leading to perpendicular heating. A common dissipation mechanism proposed for KAW-like fluctuations is Landau damping \cite[e.g.,][]{Howes2008a,Schekochihin2009}; however, this leads to heating parallel to $\textbf{B}_0$. Instead, perpendicular heating may arise from processes such as stochastic heating \citep{Chandran2010,Chandran2013} or even cyclotron resonance \citep{Isenberg}, although more work is needed to confirm this.

We note that several studies \citep{Markovskii2013a,Markovskii2016a,Markovskii2016,Vasquez2018} show that non-linear fluctuations confined to the plane perpendicular to $\mathbf{B}_0$ can produce the observed right-handed helicity signature in $\sigma_\perp$ in the same way as linear KAWs \citep[e.g.,][]{Howes2010,He2012a}. In this study, we refer to KAW-like fluctuations as non-linear turbulent fluctuations with polarisation properties that are consistent with linear KAWs, rather than linear modes. This interpretation does not preclude the possibility of resonant damping \citep{Li2016,Li2019,Klein2017,Klein2020,Howes2018,Chen2018} or stochastic heating \citep[e.g.,][]{Cerri2021} discussed above, however, additional processes cannot be ruled out. For example, kinetic simulations show perpendicular heating of ions by turbulent processes that may be unrelated to wave damping or stochastic heating, although, the exact heating mechanism is still unclear \citep[e.g.,][]{Parashar2009,Servidio2012,Vasquez2015,Yang2017}.

The variation of $T_{p,\|}/\left<T_{p,\|}\right>$ in the $\beta_p$-$\theta_{vB}$ plane in Figure \ref{fig:4}$(b)$ is more difficult to interpret. This result could also be a signature of proton Landau damping of KAW-like fluctuations, although, this process is typically stronger at $\beta_p\gtrsim1$ \citep{Gary2004a,Kawazura2019}. While we measure fluctuations that can consistently explain the enhancement in $T_{p,\|}/\left<T_{p,\|}\right>$, other fluctuations may be present that we do not measure. Direct evidence of energy transfer between the fluctuations and protons is needed to confirm this result, for example, using the field-particle correlation method \citep{Klein2016,Howes2017,Klein2017c,Klein2017b,Chen2018,Li2019}. This evidence will require higher-resolution data than provided by \textit{Wind}. We note that caution must be given when interpreting these results, since several other effects may also explain the temperature dependence seen in the $\beta_{p}$-$\theta_{vB}$ plane. For example, instrumental effects and the role of solar wind expansion may result in similar temperature profiles. We now discuss these two effects in turn, showing that they cannot fully replicate our results presented in this paper.

\section{Analysis Caveats} \label{sec:Caveats}

\subsection{Instrumentation \& Measurement Uncertainties} \label{sec:uncert}

\begin{figure}
	\centering
	\includegraphics[width=\linewidth]{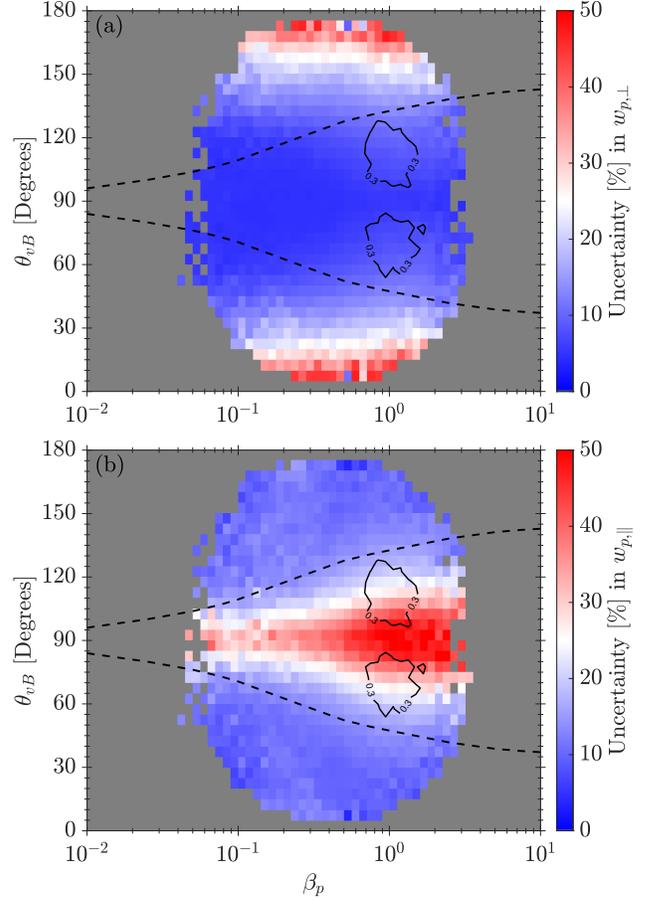} 
	\caption{Median percentage uncertainty in $(a)$ $w_{p,\perp}$ and $(b)$ $w_{p,\|}$, across $\beta_{p}$-$\theta_{vB}$ space. The dashed black lines indicate the isocontours of $\tilde{\sigma}_m(\theta_{vB})=0$ mirrored about the line $\theta_{vB}=90^\circ$. We also include contours of constant $\sigma_\perp=0.3$ from Figure \ref{fig:3}$(b)$ as black lines.}
	\label{fig:5}
\end{figure}

The SWE Faraday cups on-board \textit{Wind} measure a reduced VDF that is a function of the average direction of $\textbf{B}_0$ over the measurement interval \citep{Kasper2002}. As the spacecraft spins every 3 s in the ecliptic plane, the Faraday cups measure the current due to ions in several angular windows. The Faraday cups repeat this process using a different voltage (energy) window for each spacecraft rotation, building up a full spectrum every $\sim$92 s. By fitting a bi-Maxwellian to the reduced proton VDF, the proton thermal speeds, $w_{p,\|}$ and $w_{p,\perp}$, are obtained and converted to temperatures via $T_{p,\perp/\|}=m_pw_{p,\perp/\|}^2/2k_B$, where $m_p$ is the proton mass. Due to the orientation of the cups on the spacecraft body, the direction of $\textbf{B}_0$ with respect to the axis of the cups as they integrate over the proton VDF can cause inherent uncertainty in $w_{p,\|}$ and $w_{p,\perp}$. For example, if $\textbf{B}_0$ is radial, then measurements of $w_{p,\|}$ have a smaller uncertainty compared to when the field is perpendicular to the cup, i.e., $\textbf{B}_0$ is orientated out of the ecliptic plane by a significant angle, $\theta_{vB}\gtrsim60^\circ$ \citep{Kasper2006}. In Figure \ref{fig:5}, we plot the percentage uncertainty in $w_{p,\|}$ and $w_{p,\perp}$,

\begin{equation}
	U(w_{p,\perp/\|})=\frac{\Delta w_{p,\perp/\|}}{w_{p,\perp/\|}}\times100\%,
\end{equation}

\noindent in the $\beta_{p}$-$\theta_{vB}$ plane, where $\Delta w_{p,\perp/\|}$ is the uncertainty in $w_{p,\perp/\|}$, derived from the non-linear fitting of the distribution functions. We note that these uncertainties are not equivalent to Gaussian measurement errors; however, they provide a qualitative aid to understand systematic instrumental issues in the Faraday cup data.

We see that $w_{p,\perp}$ has a larger uncertainty ($\sim$40\%) at quasi-parallel angles in Figure \ref{fig:5}$(a)$, which is almost independent of $\beta_p$. While the median $T_{p,\perp}/\left<T_{p,\perp}\right>$ in Figure \ref{fig:4}$(a)$ is larger at these angles, it exhibits a clear dependence on $\beta_p$. Therefore, increased uncertainty in the temperature measurements alone cannot completely account for the observed enhancement in $T_{p,\perp}/\left<T_{p,\perp}\right>$ at these angles in the $\beta_{p}$-$\theta_{vB}$ plane. At quasi-perpendicular angles, the uncertainty in $w_{p,\perp}$ is less than 10\%, suggesting that the enhancements in $T_{p,\perp}/\left<T_{p,\perp}\right>$ in Figure \ref{fig:4}$(a)$ at $\beta_p\simeq1$ and $40^\circ\lesssim\theta_{vB}\lesssim140^\circ$ are unlikely to result from instrumental uncertainties. From Figure \ref{fig:5}$(b)$, the uncertainty in $w_{p,\|}$ is largest at $\theta_{vB}\simeq90^\circ$, although there is a larger spread to $70^\circ\lesssim\theta_{vB}\lesssim110^\circ$ at $\beta_p\gtrsim0.3$. By comparing with Figure \ref{fig:4}$(b)$, the enhancement in $T_{p,\|}/\left<T_{p,\|}\right>$ over the entire $\beta_p$ range does not coincide exactly with the regions of $\beta_{p}$-$\theta_{vB}$ space where these measurements have increased uncertainty. We also expect that any increased uncertainty in the $w_{p,\|}$ measurements would lead to increased noise that destroys any coherent median signal in this space, weakening the enhancement seen in Figure \ref{fig:4}$(b)$. Therefore, we conclude that the increased uncertainty in $w_{p,\|}$ at oblique angles is not the sole cause of the observed enhancement in $T_{p,\|}/\left<T_{p,\|}\right>$.

Another source of uncertainty from the SWE measurements arises from the changing magnetic field direction over the course of the $\sim$92 s measurement interval \citep{Maruca2013}. We quantify the angular fluctuations in $\textbf{B}$ using:

\begin{equation} \label{eq:devB}
	\psi_{B}=\sum_{i=1}^{N} \arccos \left(\hat{\mathbf{B}}_{i} \cdot \hat{\mathbf{B}}_{92}\right) / N,
\end{equation}

\noindent where $N$ is the number of spacecraft rotations in a single measurement, $\hat{\mathbf{B}}_{92}$ is the average magnetic field direction over the whole measurement interval, and $\hat{\mathbf{B}}_{i}$ is the magnetic field unit vector averaged over each 3 s rotation. A large $\psi_B$ can lead to the blurring of anisotropies in the proton thermal speeds. In other words, the fluctuations in $\textbf{B}$ over the integration time result in a broadening of the reduced VDFs, increasing uncertainty in these measurements \citep[e.g., see][]{Verscharen2011}. To reduce this blurring effect, we remove SWE measurements with angular deviations $\psi_B>15^\circ$. \citet{Maruca2012b} provides an alternative dataset of proton moments from SWE measurements to account for large deviations in the instantaneous magnetic field, using an average $\textbf{B}_0$ over each voltage window scan (i.e., one rotation of the spacecraft, $\sim$3 s) to calculate $w_{p,\perp}$ and $w_{p,\|}$. \citet{Maruca2013} show that the \citet{Kasper2002} dataset often underestimates the temperature anisotropy of proton VDFs. Our comparison with this alternative dataset (not shown here) reveals that both $T_{p,\perp}/\left<T_{p,\perp}\right>$ and $T_{p,\|}/\left<T_{p,\|}\right>$ show a similar, albeit slightly reduced, dependence on both $\beta_p$ and $\theta_{vB}$. This result suggests that the temperature dependence we see in the $\beta_p$-$\theta_{vB}$ plane is unlikely caused by the blurring of proton temperature anisotropy measurements.

\subsection{CGL Spherical Expansion} \label{sec:CGL}

\begin{figure}
	\centering
	\includegraphics[width=0.8\linewidth]{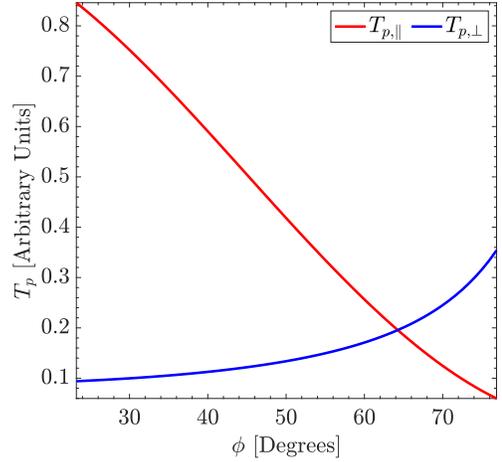} 
	\caption{The temperature profiles $T_{p,\|}$ and $T_{p,\perp}$ as functions of $\phi$ at $r\simeq1$ au for CGL spherical expansion given by Equations \ref{eq:CGL1} and \ref{eq:CGL2}.}
	\label{fig:6}
\end{figure}

Another possible source of proton temperature dependence on $\theta_{vB}$ is the expansion of the solar wind as it flows out into the heliosphere. The double adiabatic closure presented by \citet{Chew1956} predicts the evolution of $T_{p,\perp}$ and $T_{p,\|}$ assuming no collisions, negligible heat flux, and no local heating:

\begin{equation} \label{eq:CGL}
	\frac{d}{dt}\left(\frac{T_{p,\perp}}{B}\right)=0 \quad \text { and } \quad \frac{d}{dt}\left(\frac{T_{p,\|} B^{2}}{n_{p}^{2}}\right)=0,
\end{equation}

\noindent where $d/dt$ is the convective derivative. Under the assumption of steady-state spherical expansion, which is purely transverse to the radial direction with a constant radial velocity, $\textbf{v}_{sw}=v_{sw}\,\hat{\textbf{R}}$, the continuity equation gives $n_p\propto1/r^2$, where $r$ is the radial distance from the Sun. We assume that the radial evolution of the magnetic field in the equatorial plane follows the Parker spiral \citep{Parker1958},

\begin{equation} \label{eq:ParkerSpiral}
	B \propto \frac{\sqrt{\cos^2{\phi_{0}}+r^2\sin^2{\phi_{0}}}}{r^{2}},
\end{equation}

\noindent which gives a radial dependence of $B\propto1/r^2$ when $\phi_0=0^\circ$ and $B\propto1/r$ when $\phi_0=90^\circ$. The angle, $\phi_0$, is the foot-point longitude of the field at the solar wind source surface, given by:

\begin{equation} \label{eq:phi0}
	\phi_0=\arctan\left(\frac{r_0\Omega_{\sun}}{v_{sw}}\right),
\end{equation}

\noindent where $\Omega_{\sun}=2.85\times10^{-6}$ rad/s is the constant solar angular rotation rate and $r_0\simeq20R_{\sun}$ \citep{Owens2013}. Therefore, a value of $v_{sw}$ sets the value of $\phi_0$ at a given radius, $r_0$. Then, $\phi=\tan^{-1}{(B_\phi/B_r)}$ is the azimuthal angle of $\textbf{B}$ in the equatorial plane at a distance, $r$, from the Sun. The two angles are related by $\tan{\phi}=R\tan{\phi_0}$, where $R=r/r_0$. From Equations \ref{eq:CGL}, \ref{eq:ParkerSpiral}, and the radial dependence of $n_p$, we obtain:

\begin{equation}
	\frac{T_{p,\perp}}{T_{0,\perp}}=\frac{\sqrt{\cos ^{2}\left(\phi_{0}\right)+R^{2} \sin ^{2}\left(\phi_{0}\right)}}{R^{2}}, \label{eq:CGL1}
\end{equation}\\

\noindent and

\begin{equation}
	\frac{T_{p,\|}}{T_{0,\|}}=\frac{1}{\cos ^{2}\left(\phi_{0}\right)+R^{2} \sin ^{2}\left(\phi_{0}\right)}, \label{eq:CGL2}
\end{equation}

\noindent where $T_{0,\perp}$ and $T_{0,\|}$ are the perpendicular and parallel proton temperatures at $r_0$, respectively. We use Equations \ref{eq:CGL1} and \ref{eq:CGL2} to investigate the dependence of proton temperature on $\phi$ at $r=215R_{\sun}\simeq1$ au. Since the solar wind velocity is radial, the angle $\phi$ is approximately equal to $\theta_{vB}$. We set $T_{0,\perp}=10$ and $T_{0,\|}=1$, giving $R=10.75$ for $r_0=20R_{\sun}$. We create a distribution of angles $\phi_0$ using Equation \ref{eq:phi0} by selecting a range of wind speeds: $100\leq v_{sw}\leq 1000$ km/s. This range of $\phi_0$ gives $20^{\circ} \lesssim \phi \lesssim 80^{\circ}$ at 1 au. In Figure \ref{fig:6}, we show the variation of $T_{p,\perp}$ and $T_{p,\|}$ with $\phi$. We choose a larger $T_{\perp,0}$ to show more clearly the variation in $T_{p,\perp}$. We see that $T_{p,\|}$ remains similar to the value set close to the Sun (small $\phi$), and decreases rapidly with increasing $\phi$, approaching zero at $\phi\gtrsim70^\circ$. On the other hand, $T_{p,\perp}$ is largest at $\phi\gtrsim60^\circ$ and approaches 0.1 for $\phi\lesssim30^\circ$. This dependence of $T_{p,\perp}$ and $T_{p,\|}$ is opposite to what we observe in Figure \ref{fig:4}, which in general shows larger $T_{p,\|}$ at $\theta_{vB}\simeq90^\circ$ and larger $T_{p,\perp}$ at $\theta_{vB}\simeq0^\circ$. Therefore, spherical expansion alone cannot explain our results.

\section{Conclusions} \label{sec:Conclusions}

We use magnetic helicity to investigate the polarisation properties of Alfvénic fluctuations with finite radial wave-number, $k_r$, at ion-kinetic scales in the solar wind. Using almost 15 years of \textit{Wind} observations, we separate the contributions to helicity from fluctuations with wave-vectors quasi-parallel and oblique to $\mathbf{B}_0$, finding that the helicity of Alfv\'enic fluctuations is consistent with predictions from linear Vlasov theory. In particular, the peak in magnetic helicity at ion-kinetic scales and its variation with $\beta_p$ and $\theta_{vB}$ shown in Figure \ref{fig:3} are in agreement with the dispersion relation of linear Alfv\'en waves \citep{Gary1986}, when modified by Taylor's hypothesis. This result suggests that the non-linear turbulent fluctuations at these scales share at least some polarisation properties with Alfv\'en waves.

We also investigate the dependence of local kinetic heating processes due to turbulent dissipation on $\theta_{vB}$. In Figure \ref{fig:4}, we find that both $T_{p,\perp}$ and $T_{p,\|}$, when normalised to their average value in each $\beta_p$-bin, show a clear dependence on $\theta_{vB}$. The temperature parallel to $\mathbf{B}_0$ is generally higher in the parameter-space where we measure a coherent helicity signature associated with KAW-like fluctuations, and perpendicular temperature higher in the parameter-space where we measure a signature expected from AIC waves. We also see small enhancements in the perpendicular temperature where we measure the strongest helicity signal of KAW-like fluctuations. However, we re-iterate the important fact that the lack of a wave signal is not the same as a lack of presence of waves.

Our results suggest that the properties of turbulent fluctuations at ion-kinetic scales in the solar wind depends on the angle $\theta_{vB}$. This finding is inconsistent with the general assumption that sampling different $\theta_{vB}$ allows us to sample different parts of the same ensemble of fluctuations that is otherwise unaltered in its statistical properties. Therefore, studies that sample different $\theta_{vB}$ in order to sample different $\theta_{kB}$ need to be interpreted very carefully. Instead, if we assume that the dissipation mechanisms and proton heating depend on $\theta_{vB}$, the enhancements in proton temperature in Figure \ref{fig:4} are consistent with the role of wave-particle interactions in determining proton temperature in the solar wind. For example, whenever we measure the helicity of AIC waves or KAW fluctuations, then we also measure enhancements in proton temperature. However, the inverse is not necessarily true. We suggest that heating mechanisms associated with KAWs lead to both parallel \citep{Howes2008a,Schekochihin2009} and perpendicular \citep{Chandran2010,Chandran2013,Isenberg} heating. We rule out both instrumental and large-scale expansion effects, finding that neither of them alone can explain the observed temperature profile in the $\beta_ p$-$\theta_{vB}$ plane.

In summary, our observations suggest that the properties of Alfv\'enic fluctuations at ion-kinetic scales determine the level of proton heating from turbulent dissipation. This interpretation is consistent with recent studies showing that larger magnetic helicity signatures at ion-kinetic scales are associated with larger proton temperatures and steeper spectral exponents \citep{Pine2020c,Zhao2020d,Zhao2021}. Our findings, therefore, provide new evidence for the importance of local kinetic processes in determining proton temperature in the solar wind. We emphasise that our conclusions do not invoke causality, just correlation. For example, we cannot rule out a lack of cooling rather than heating. However, while the adiabatic expansion of the solar wind causes the temperature to vary with $\theta_{vB}$, this cannot explain the observed temperature profiles in the $\beta_p$-$\theta_{vB}$ plane. Further work is ongoing in order to confirm these results and develop a theory for the processes associated with the polarisation properties of Alfv\'enic fluctuations that lead to the observed temperature profiles.

\begin{acknowledgments}
	LDW thanks B. Alterman, L. Matteini, D. Stansby, and J. Stawarz for useful comments and discussions. LDW is also grateful to B. Maruca for sharing his processed SWE dataset for validation testing. LDW was supported by an STFC studentship at UCL/MSSL ST/N504488/1 and the STFC consolidated grant ST/S000364/1 to Imperial College London; DV was supported by STFC Ernest Rutherford Fellowship ST/P003826/1; RTW was supported by the STFC consolidated grants to UCL/MSSL, ST/N000722/1 and ST/S000240/1. JMT was supported by NSF SHINE award (AGS-1622306). GGH was supported by NASA grants 80NSSC18K1217 and 80NSSC18K0643. All data from the \textit{Wind} spacecraft used in this study are publicly available and were obtained from the NASA \href{http://spdf.gsfc.nasa.gov}{SPDF web-site}. The NHDS code is available at \href{https://github.com/danielver02/NHDS}{https://github.com/danielver02/NHDS}.
\end{acknowledgments}

\appendix

\section{Decomposition of Fluctuating Magnetic Helicity}

\begin{figure}
	\centering
	\includegraphics[width=\linewidth]{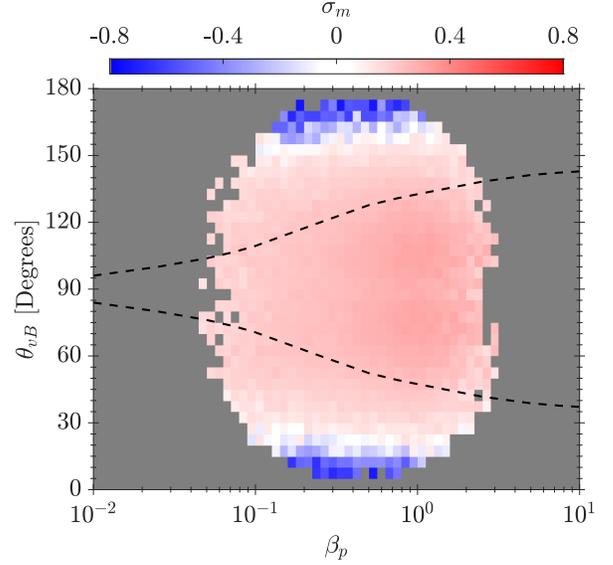} 
	\caption{Median value of the peak in $\sigma_m^r(k_r)$ across the $\beta_p$-$\theta_{vB}$ plane. The dashed black lines indicate the isocontours of $\tilde{\sigma}_m(\theta_{vB})=0$ mirrored about the line $\theta_{vB}=90^\circ$.}
	\label{fig:A2}
\end{figure}

Here we present a mathematical derivation that decomposes $\sigma^r_m(k_r)$ into the different contributions, $\sigma_{ij}(k_l)$, calculated using Equation \ref{eqnhelFA} \citep[see also][]{Wicks2012}. In Figure \ref{fig:A2}, we plot the median value of the peak in $\sigma_m^r(k_r)$ across the $\beta_p$-$\theta_{vB}$ plane, showing two helicity signatures of opposite handedness. This technique allows us to separate the helicity signatures of different fluctuations at ion-kinetic scales in the solar wind, as we show in Figure \ref{fig:3}.

We consider a spacecraft that samples a single mode with wave-vector:

\begin{equation}
	\mathbf{k} = k_\perp \cos \alpha \, \hat{\mathbf{x}}   +  k_\perp \sin \alpha \, \hat{\mathbf{y}} + k_\parallel \, \hat{\mathbf{z}},
\end{equation}

\noindent where $\alpha$ is the azimuthal angle of $\textbf{k}$ in the $x$-$y$ plane. The full signal from turbulence corresponds to a superposition of the signals from each of the modes, so considering a single mode is sufficient to understand how the components $\sigma_{ij}(k_l)$ are related to $\sigma^r_m(k_r)$. Without loss of generality, we take the solar wind velocity to be in the $x$-$z$ plane,

\begin{equation}
	\mathbf{v}_{sw} = v_{sw} \sin \theta_{vB} \, \hat{\mathbf{x}} +  v_{sw} \cos \theta_{vB} \, \hat{\mathbf{z}},
\end{equation}

\noindent and the local mean magnetic field to be $\mathbf{B}_0=B_0 \, \hat{\mathbf{z}}$. We use the relation $2\,\mathrm{Im}\left\lbrace a^*b\right\rbrace\equiv i\,(ab^* - a^*b)$ to rewrite $H'^r_m(k_r)$ in RTN coordinates from Equation \ref{eq:redflucthel} into the form:

\begin{equation}
	H'^r_m(k_r)=i \frac{\delta B_T \delta B^*_N-\delta B^*_T \delta B_N}{k_r}.
\end{equation}

\noindent The normalised reduced fluctuating magnetic helicity, $\sigma^r_m(k_r)$, is then given by Equation \ref{eq:reducednormhel}. We define the relation between the RTN and the field-aligned (Equation \ref{eqncoords}) coordinate systems using the unit vector along the sampling direction, $\hat{\mathbf{v}}_{sw}=\mathbf{v}_{sw}/|\mathbf{v}_{sw}|$:

\begin{eqnarray}
	\hat{\mathbf{R}} &=& \hat{\mathbf{v}}_{sw} = \sin \theta_{vB} \hat{\mathbf{x}}
	+ \cos \theta_{vB} \hat{\mathbf{z}};\nonumber\\
	\hat{\mathbf{T}} &=& \hat{\mathbf{B}}_0 \times \hat{\mathbf{v}}_{sw}  /| \hat{\mathbf{B}}_0 \times\hat{\mathbf{v}}_{sw} | =  
	\hat{\mathbf{y}};\\
	\hat{\mathbf{N}} &=& \hat{\mathbf{R}} \times \hat{\mathbf{T}} = -\cos \theta_{vB}  \hat{\mathbf{x}} + \sin \theta_{vB} \hat{\mathbf{z}}.\nonumber
\end{eqnarray}

\noindent By substituting for the RTN unit vectors in terms of $\hat{\mathbf{x}}$, $\hat{\mathbf{y}}$, and $\hat{\mathbf{z}}$ and simplifying, we obtain:

\begin{eqnarray}
	\sigma^r_m(k_r)&=&\frac{1}{|\delta\mathbf{B}(k_r)|^2} \biggl\{
	\left[ \frac{ i (\delta B_x \delta B^{*}_y-\delta B^{*}_x \delta B_y)}{k_z} \right]k_z\cos \theta_{vB} \nonumber\\
	&&+
	\left[ \frac{ i (\delta B_x \delta B^*_z-\delta B^*_x \delta B_z)}{k_y} \right]k_y  \sin \theta_{vB} \sin\alpha \nonumber\\
	&&+
	\left[ \frac{ i (\delta B_y \delta B^*_z-\delta B^*_y \delta B_z)}{k_x} \right]k_x  \sin \theta_{vB}  \cos \alpha \biggl\}.\nonumber\\
	\label{eq:sigmar_trans}
\end{eqnarray}

\noindent By defining the non-reduced fluctuating magnetic helicity (see also, Equation \ref{eq:NRhel}) as:

\begin{eqnarray}
	H'_m(\mathbf{k})&=&i \frac{\delta B_x \delta B^*_y-\delta B^*_x \delta B_y}{k_z} \nonumber\\
	&\equiv& i \frac{\delta B_y \delta B^*_z-\delta B^*_y \delta B_z}{k_x}
	\equiv i \frac{\delta B_z \delta B^*_x-\delta B^*_z \delta B_x}{k_y}, \nonumber\\
\end{eqnarray}

\noindent we can manipulate Equation \ref{eq:sigmar_trans} into the form,

\begin{eqnarray}
	\sigma^r_m(k_r)&=&\frac{H'_m(\mathbf{k})}{|\delta\mathbf{B}(k_r)|^2}\bigl\{
	k_z\cos \theta_{vB} -k_y  \sin \theta_{vB} \sin \alpha \nonumber\\
	&&+
	k_x\sin \theta_{vB}  \cos \alpha \bigl\},\nonumber\\\nonumber\\
	&\equiv&\sigma_{xy}(k_z) \cos \theta_{vB} +  \sigma_{xz}(k_y)  \sin \theta_{vB} \sin \alpha \nonumber\\
	&&+
	\sigma_{yz}(k_x) \sin \theta_{vB}  \cos \alpha,
	\label{eq:sigmar_trans2}
\end{eqnarray}

\noindent where we define the different contributions, $\sigma_{ij}(k_l)$, using Equation \ref{eqnhelFA}. We equate each of the terms between the two forms in Equation \ref{eq:sigmar_trans2} to obtain the following direct relations between $\sigma^r_m(k_r)$ and $\sigma_{xy}(k_z)$,  $\sigma_{xz}(k_y)$ and $\sigma_{yz}(k_x)$:

\begin{equation}
	\sigma_{xy}(k_z) = \frac{H'_m(\mathbf{k})}{|\delta\mathbf{B}(k_r)|^2} k_z = \sigma^r_m(k_r) \frac{k_z}{k_r},
\end{equation}

\begin{equation}
	\sigma_{xz}(k_y) = -\frac{H'_m(\mathbf{k})}{|\delta\mathbf{B}(k_r)|^2} k_y = -\sigma^r_m(k_r) \frac{k_y}{k_r},
\end{equation}

\noindent and,

\begin{equation}
	\sigma_{yz}(k_x) = \frac{H'_m(\mathbf{k})}{|\delta\mathbf{B}(k_r)|^2} k_x = \sigma^r_m(k_r) \frac{k_x}{k_r},
\end{equation}

\noindent which are the same as Equations \ref{Eq:a1}, \ref{Eq:a2}, and \ref{Eq:a3}.

\begin{figure}
	\centering
	\includegraphics[width=0.95\linewidth]{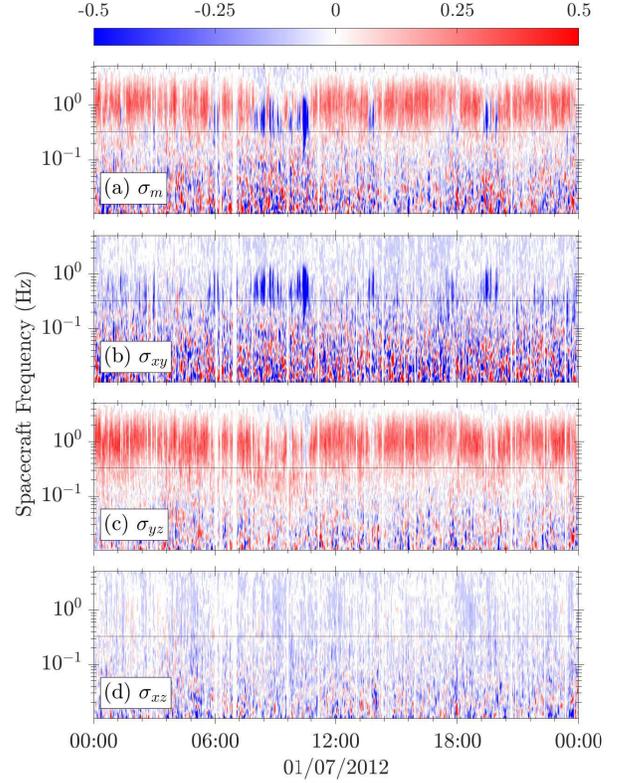} 
	\caption{$(a)$ Time series of reduced normalised fluctuating magnetic helicity spectra, $\sigma^r_m$, for a day of observations on 01/07/2012. The different contributions to the total helicity, $(b)$ $\sigma_{xy}$, $(c)$ $\sigma_{yz}$, and $(d)$ $\sigma_{xz}$, respectively. We plot the spectra as functions of the frequency in the spacecraft frame, $f_{sc}=\omega_{sc}/2\pi$. The black dashed line in panel $(d)$ is the spacecraft spin frequency, 0.33 Hz.}
	\label{fig:A1}
\end{figure}

To highlight the separation of different fluctuations in the solar wind using this technique, we show in Figure \ref{fig:A1}$(a)$ a time series of magnetic helicity spectra, $\sigma_m^r$, measured by \textit{Wind} on 01/07/2012. We plot the spectra as functions of frequency in the spacecraft frame, $f_{sc}=\omega_{sc}/2\pi$ (see Equation \ref{eq:Taylor}). In panels $(b)$-$(d)$, we also plot $\sigma_{ij}$, showing the decomposition of $\sigma_m^r$ into its three components. The two coherent signatures of opposite handedness at $f_{sc}\simeq1$ Hz in panel $(a)$ are completely separated into the components $\sigma_{xy}$ and $\sigma_{yz}$ in panels $(b)$ and $(c)$. In panel $(d)$, we see only small enhancements close to 0.33 Hz, which corresponds to the spin frequency of the spacecraft. Besides this spacecraft artefact, there is no coherent helicity signature in $\sigma_{xz}$, as expected.

\bibliographystyle{yahapj}
\bibliography{Bibliography}

\end{document}